\documentclass[epjCONF]{svjour}
\usepackage{graphics}
\ifx\pdfoutput\undefined
  \usepackage{graphicx}
\else
  \usepackage[pdftex]{graphicx}
  \usepackage{epstopdf}
\fi
\usepackage[varg]{txfonts} 
\usepackage[latin1]{inputenc}
\session-title{MESON2012}
\begin{document}
\title{ Hadron Spectroscopy with COMPASS -- Newest Results}
\author{Frank Nerling\thanks{\email{nerling@cern.ch}}, for the COMPASS collaboration}
\institute{Physikalisches Institut, Albert-Ludwigs-Universit\"at Freiburg / CERN PH Department, Geneva}
\abstract{
The COMPASS experiment at the CERN SPS investigates the structure and spectrum of hadrons by 
scattering high energetic hadrons and polarised muons off various fixed targets. During the years 
2002-2007, COMPASS focused on nucleon spin physics using 160 GeV/c polarised $\mu^+$ beams on 
polarised deuteron and proton targets, including measurements of the gluon contribution to the 
nucleon spin using longitudinal target polarisation as well as studies of transverse spin effects 
in the nucleon on a transversely polarised target. One major goal of the physics programme using 
hadron beams is the search for new states, in particular the search for $J^{PC}$ exotic states and 
glue-balls. COMPASS measures not only charged but also neutral final-state particles, allowing for 
investigation of new objects in different reactions and decay channels. In addition COMPASS can 
measure low-energy QCD constants like, e.g. the electromagnetic polarisability of the pion. Apart 
from a few days pilot run data taken in 2004 with a 190 GeV/c $\pi^{-}$ beam on a Pb target, showing 
a significant spin-exotic $J^{PC}$ = $1^{-+}$ resonance at around 1660 MeV/$c^{2}$, COMPASS collected 
high statistics with negative and positive 190 GeV/$c$ hadron beams on a proton (H$_2$) and nuclear 
(Ni, Pb) targets in 2008 and 2009. We give a selected overview of the newest results and discuss the 
status of various ongoing analyses.
} 
\maketitle
\section{Introduction}
\label{sec.intro}
The COMPASS fixed-target experiment is a facility to study QCD, covering a large range in momentum 
transfer from larger than 1\,GeV$^{2}$/$c^{2}$ down to smaller than 10$^{-3}$\,GeV$^{2}$/$c^{2}$. 
Physics topics addressed is the study of hadron structure and dynamics, comprising the investigation 
of nucleon spin-structure using a polarised muon beam, see e.g.~\cite{badelek:2012} for a recent summary, 
and hadron spectroscopy using hadron beams to study the mass spectrum of hadrons, to test Chiral Perturbation 
Theory (ChPT) and to measure the fundamental quantities of pion and kaon polarisibilities.  

One important goal of the physics programme using hadron beams is the search for new states, 
like spin-exotic mesons and glue-balls. Such states are beyond the simple Constituent Quark Model (CQM), however, 
they are allowed and even predicted within Quantum Chromodynamics (QCD). Their existence has been speculated 
about almost since the introduction of colour~\cite{Jaffe:1976,Barnes:1983} and the experimental proof would be 
a fundamental confirmation of QCD. Even though several candidates have been reported in the past by various 
experiments, especially their resonance nature is still highly disputed in the community, a recent overview is 
given by~\cite{MeyerHaarlem:2010}.

As a first input to the puzzle, COMPASS observed a significant $J^{PC}$ spin-exotic signal in the 2004 
pilot run data (190\,GeV/$c$ $\pi^{-}$ beam, Pb target) in three charged pion final states consistent with 
the disputed $\pi_1(1600)$~\cite{Alekseev:2009a}. 
With the high statistics data COMPASS collected in 2008 using a 190 GeV/$c$ negative pion beam scattered off 
a liquid hydrogen (proton) target, we have access to all relevant decay channels in which hybrid candidates were 
reported in the past, see e.g.~\cite{nerling:2011b}. Especially the fact that COMPASS measures not only charged 
final state particles but also neutral particles, like $\pi^{0}$, $\eta$, and $\eta'$, is of great 
advantage, allowing for independent confirmation of new states within the same experiment. 
\section{Results on Hadron Spectroscopy}
\label{sec.1}
In this section, we summarise the current status and newest results from partial-wave analyses (PWA) of the 2004 pilot 
run and the 2008 data with the focus on the search for the $\pi_1(1600)$ resonance with exotic $J^{PC}=1^{-+}$ 
quantum numbers in the $\rho\pi$, $\eta\pi$ and $\eta'\pi$, and $f_1\pi$ decay channels, moreover, we briefly discuss further 
selected analyses.
\vspace{-0.3cm}
\subsection{Diffractively produced $(3\,\pi)^{-}$ final states}
\label{sec.1.1}
\begin{figure}[tp!]
  \begin{minipage}[h]{.48\textwidth}
    \begin{center}
      \vspace{-0.5cm}
\resizebox{0.9\columnwidth}{!}{%
     \includegraphics[clip,trim= 5 0 10 15, width=0.7\linewidth, angle=0]{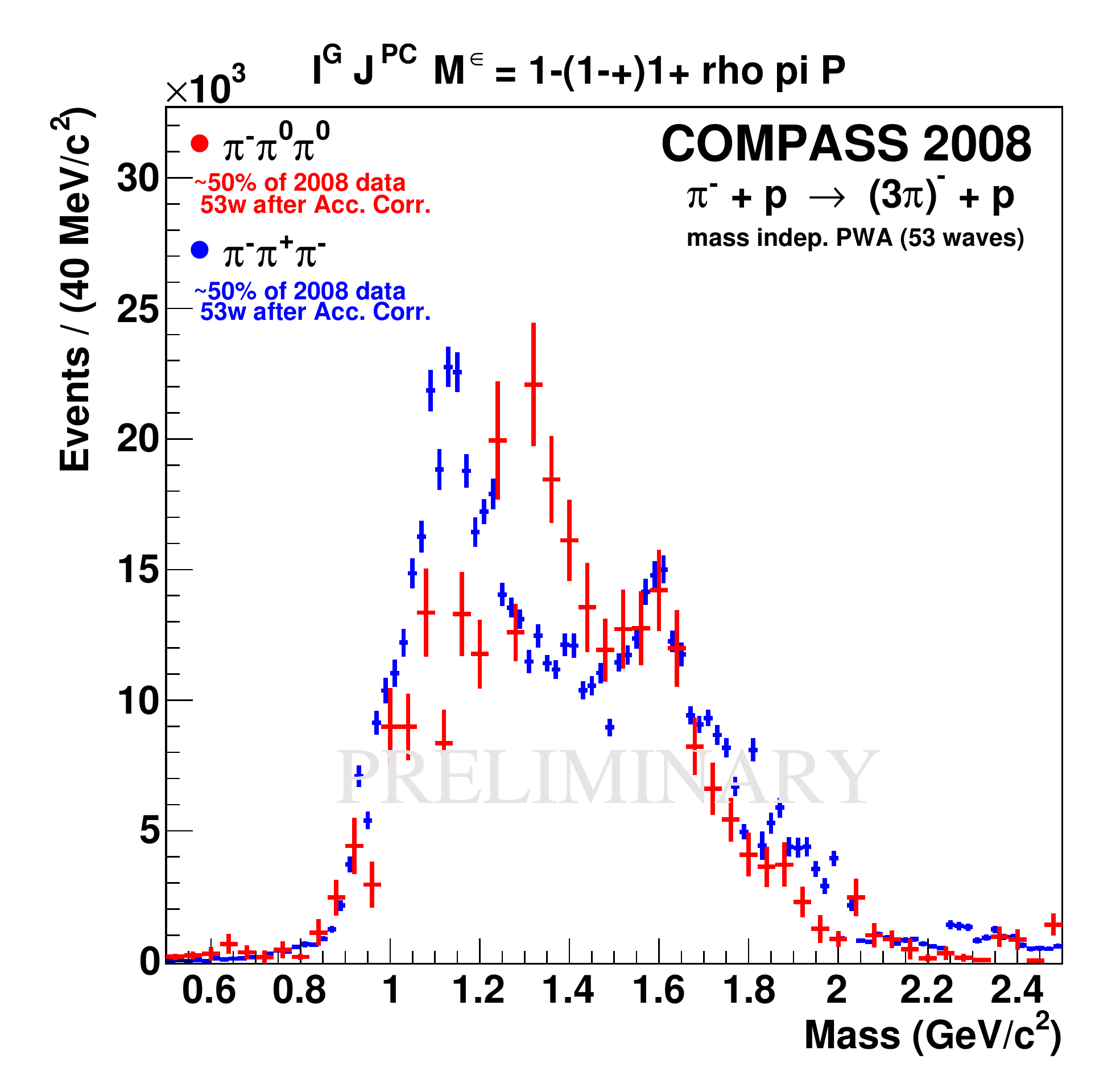} }
    \end{center}
  \end{minipage}
  \hfill
  \begin{minipage}[h]{.48\textwidth}
    \begin{center}
      \vspace{-0.5cm}
\resizebox{1.0\columnwidth}{!}{%
  \includegraphics[clip,trim= 5 0 10 15, width=1.0\linewidth, angle=0]{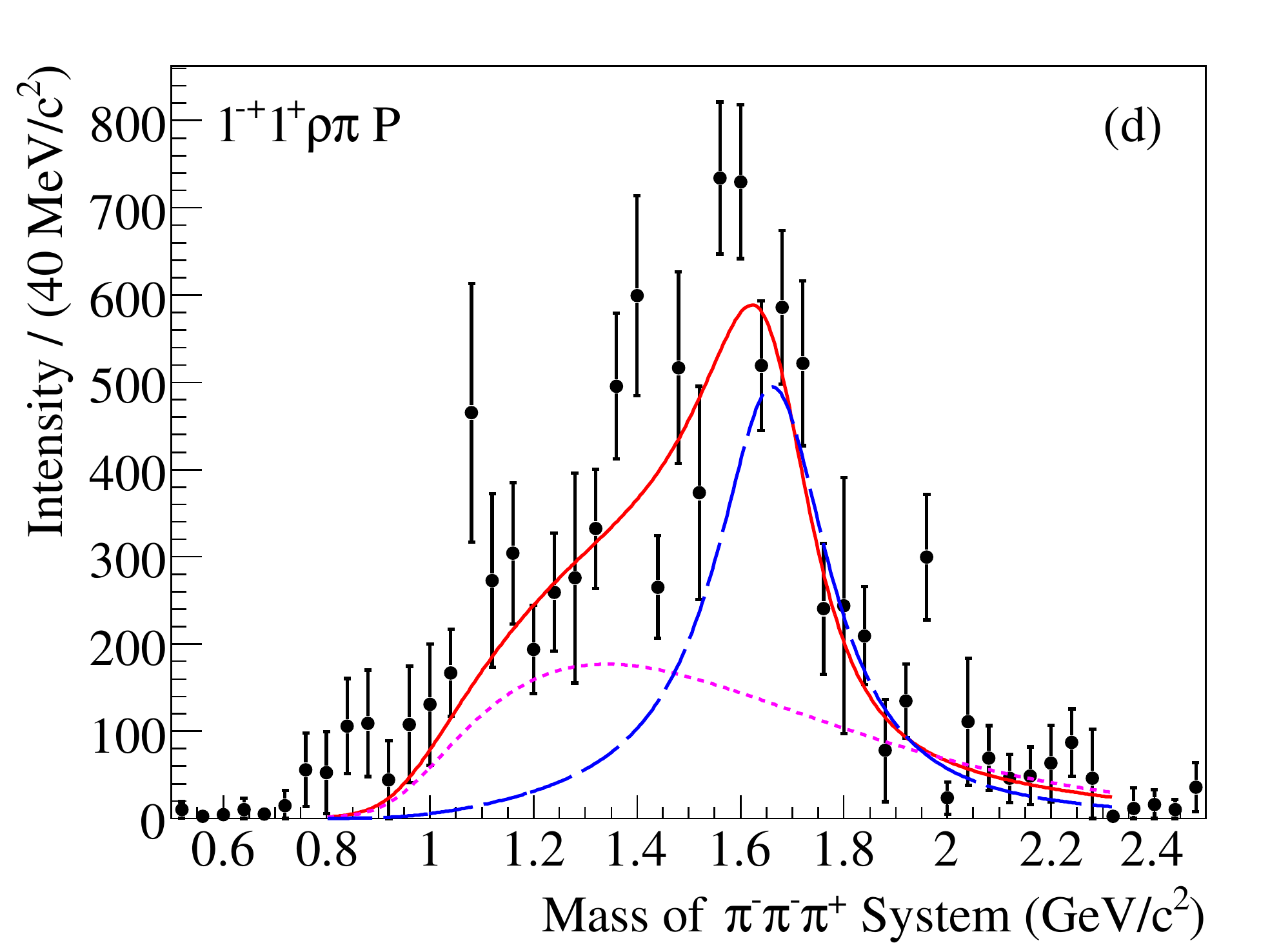} }
    \end{center}
  \end{minipage}
  \begin{minipage}[h]{.48\textwidth}
    \begin{center}
      \vspace{-0.6cm}
\resizebox{0.9\columnwidth}{!}{%
     \includegraphics[clip,trim= 5 0 10 15, width=0.7\linewidth, angle=0]{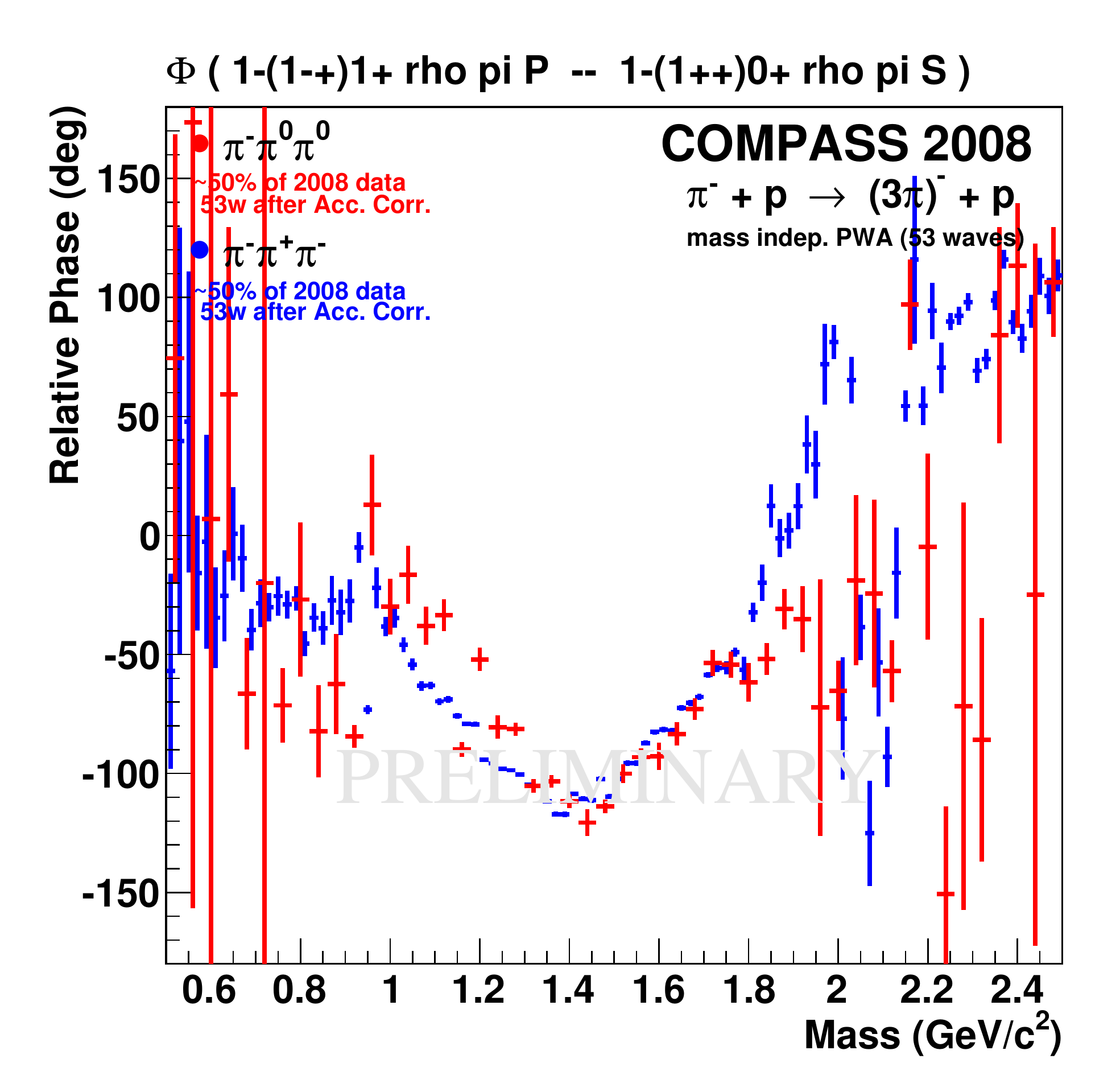} }
    \end{center}
  \end{minipage}
  \hfill
  \begin{minipage}[h]{.48\textwidth}
    \begin{center}
      \vspace{-0.5cm}
\resizebox{1.0\columnwidth}{!}{%
  \includegraphics{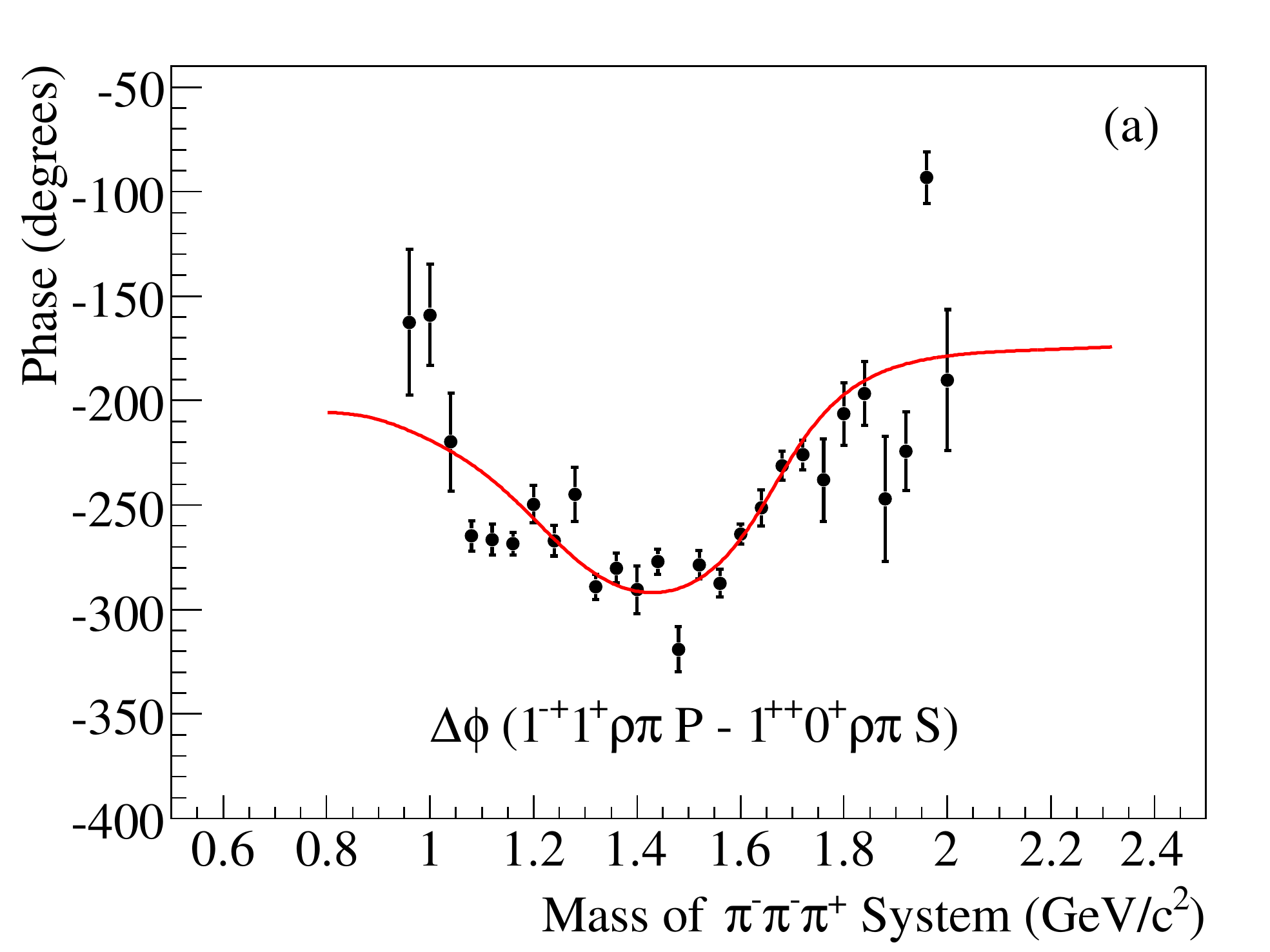} }
    \end{center}
  \end{minipage}
    \begin{center}
     \vspace{-0.3cm}
     \caption{Mass-independent PWA result for the exotic $1^{-+}$ wave in $(3\pi)^{-}$ final states for the 2008 data {\it (left)}.
       The fitted intensity of $(1^{-+})1^+\,\rho\pi\,P$ wave is shown for the neutral mode data in comparison to 
       the charged one {\it (top, left)}, the relative phase difference with respect to the $(1^{++})0^+\,\rho\pi\,S$ 
       wave for both decay modes is shown as well {\it (bottom, left)}. The published result on three charged pion 
       final states from the 2004 pilot run data~\cite{Alekseev:2009a} is shown for comparison {\it (right)}.}  
        \label{fig:exotic}
     \end{center}
     \vspace{-0.8cm}
\end{figure}

\vspace{-0.3cm}
The present mass-independent PWA results of the search for the $\pi_1(1600)$ in the $\rho\pi$ decay channel based on 
the 2008 data is compared for the neutral and the charged $3\pi$ decay modes in Fig.\,\ref{fig:exotic}, left. The PWA 
model applied is essentially the same as it was used for the published result (mass-independent and mass-dependent fits 
overlayed) that is is given for comparison (Fig.\,\ref{fig:exotic}, right), a short detailed description of the two step 
PWA method can be found in {\it e.g.}~\cite{nerling:2009}.  
%
%

In the mass-independently fitted intensities (Fig.\,\ref{fig:exotic}, left/top), two features appear for the 
neutral and charged data on top of a relatively large (presumably non-resonant, Deck-like) background. A larger peak appears 
for the neutral and the charged mode results at about 1.3\,GeV/$c^2$ and about 1.1\,GeV/$c^{2}$, respectively, which 
are still subject of detailed systematic studies (dedicated studies of background from Deck, leakage). 
Secondly, we find a smaller object at about 1.6\,GeV/$c^2$ that is consistently observed in the neutral and charged 
mode results, just in the mass region where previous experiments reported the spin-exotic $\pi_1(1600)$ resonance.

In the corresponding phase differences ${\rm \Phi}$ between the $\pi_1(1600)$ candidate and the $a_1(1260)$
shown for neutral and charged mode PWA results in Fig.\,\ref{fig:exotic}, left/bottom, a clean, rapid variation of 
${\rm \Phi}$ is observed exactly in the mass region of about $1.4-1.8\,$GeV/$c^2$. The phase differences are consistently 
observed coinciding for both $(3\pi)^{-}$ decay modes. This strongly indicates the object present at about 1.6\,GeV/$c^2$ 
being of resonant nature. Backgrounds appear differently in the phase differences below 1.4\,GeV/$c^2$ as in the fitted 
intensities. A dedicated, more detailed discussion of the comparison of the present neutral and charged mode 
PWA results is given by~\cite{nerling:2011,nerling:2012}. 

Comparing these new results from the proton target data (mass-dependent fit of the Breit-Wigner description not yet 
released) to the spin-exotic resonance found in the 2004 Pb target data at $1660$$\pm$$10^{+0}_{-64}$\, MeV/c$^2$~\cite{Alekseev:2009a}, 
they appear rather consistent. That the exotic signal at 1.6\,GeV/$c^{2}$ produced on a proton target (Fig.\,\ref{fig:exotic}, left/top) is smaller 
than the one produced on Pb (relatively compared to the non-resonant background beneath, which shows the same shape not 
only for both decay modes discussed here, but also a similar one for the 2004 Pb target data result), is qualitatively and quantitatively 
understood to be a systematic difference in the production of $M$=$1$ versus $M$=$0$ waves.
A significant dependence on the target material for the production strengths of partial-waves with spin projection $M$=$0$ versus $M$=$1$ 
has been found comparing the $(3\pi)^{-}$ PWA results for the 2004 Pb (charged) and the 2008 (charged and neutral) proton target data. 
The production strength of $M$=$1$ waves is observed to be suppressed on a proton as compared to the heavier nuclear Pb target and vice 
versa for $M$=$0$ waves, whereas the corresponding spin totals (fitted intensities summed up for both spin projections $M$), do not show 
any significant difference. These observations are re-confirmed in the 2009 Pb target data (charged mode)~\cite{haas:2009,haas:2011}, 
again in comparison to the 2008 proton target data results (to compensate for the different statistics collected and analysed, all 
those comparisons were consistently done after normalisation to the well-known and narrow $a_2(1320)$ resonance).

The state at 1.6\,GeV/$c^{2}$ observed in the spin-exotic wave with $M$=$1$ produced on proton discussed here, indicated to be of resonant
nature by the observed rapid phase motions (Fig.\,\ref{fig:exotic}, left), is consistently found smaller than the one produced on the
Pb data (charged mode)~\cite{Alekseev:2009a} (Fig.\,\ref{fig:exotic}, right).
\vspace{-0.3cm}
\subsection{Diffractively produced $(5\,\pi)^{-}$ final states}
\label{sec:1.2}
\vspace{-0.3cm}
\begin{figure}[tp!]
    \begin{center}
      \vspace{-0.2cm}
      \includegraphics[clip,trim= 0 0 0 0,width=0.75\linewidth]
      {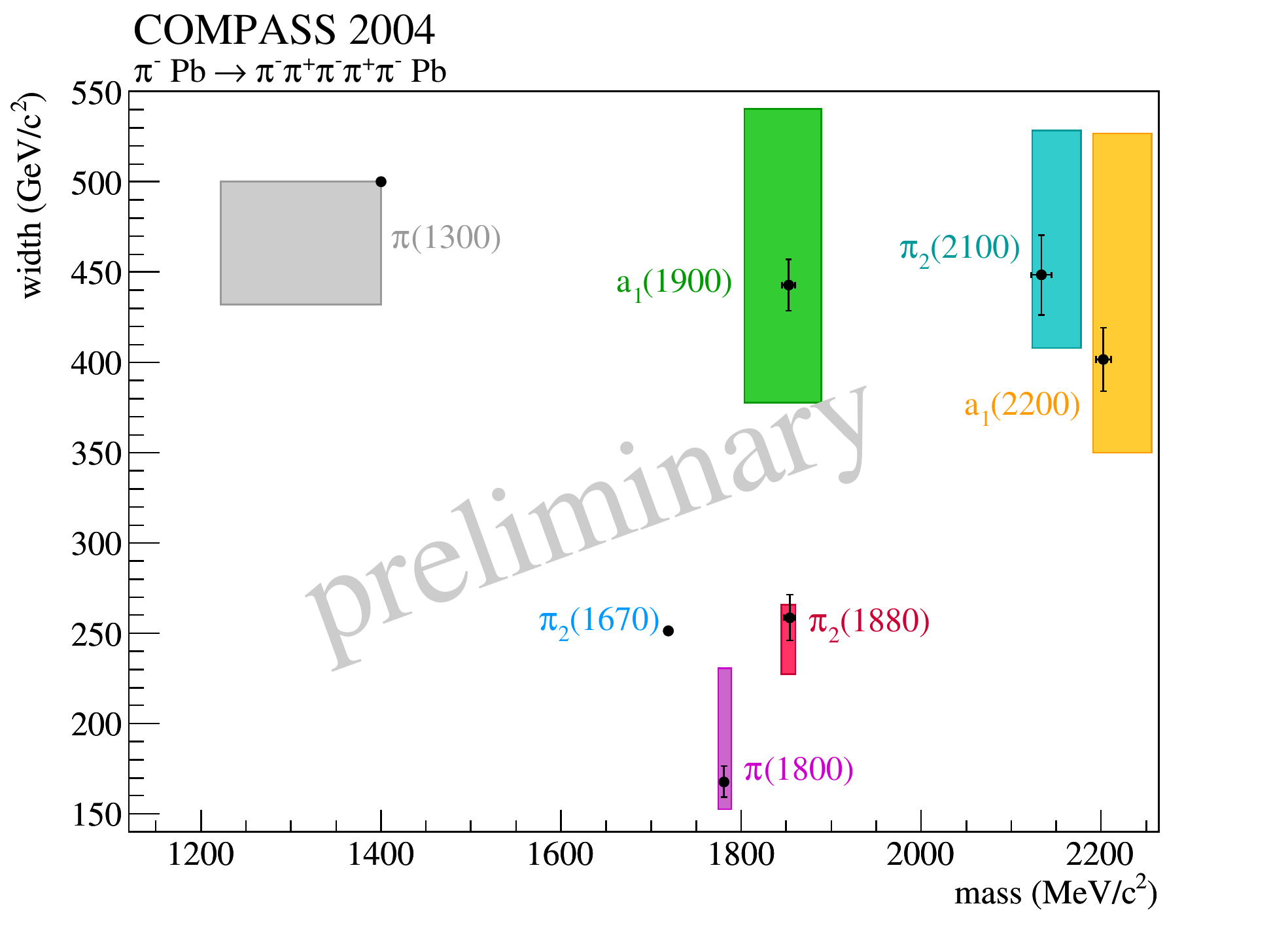}
      \caption{Extracted resonance parameters for the mass-dependent Breit-Wigner fit comprising seven resonances. The black points 
	are the best fitting values, the errors bars give the statistical uncertainties and the systematic uncertainties are given 
	by the coloured boxes, including effects of different fitting ranges, using or neglecting the covariances between real and 
	imaginary parts of the interferences terms as well as the choice of wave-set.}
      \label{fig:5piResults}
      \vspace{-0.3cm}
    \end{center}
     \vspace{-0.5cm}
\end{figure}
The studies of the $3\pi$ system in the 2004 and 2008 data, cf. Sec.\,\ref{sec.1.1}, prove the excellent acceptance not only for charged tracks but 
also for neutral particle detection. Going to the $5\pi$ system is the natural step. As charge conversion reactions at COMPASS energies 
(190\,GeV/$c$ beam momentum) are strongly suppressed, a final state produced in diffractive pion dissociation is also negatively charged.
Any $4\pi$ final state therefore contains at least one $\pi^{0}$. For the 2004 data limited in statistics (especially for final states involving 
neutral particles like $\pi^{0}$ due to the significantly lower detection efficiencies), the next simple fully charged final state populated 
contains five charged pions. Even though we do not expect to find spin-exotic $5\pi$ resonances in the low momentum transfer range of $0 < t' < 0.005$ 
analysed from the 2004 data~\cite{neubert:2009} discussed here, the PWA results on diffractive dissociation into $5\pi$ final states allow for 
precise study of various resonances not only in the light meson sector, but also exceeding 2\,GeV/$c^{2}$. The present status of determined mass and 
width parameters are shown in Fig.\,\ref{fig:5piResults}. These results are also under preparation to be published soon.   
\vspace{-0.3cm}
\subsection{Diffractively produced $\eta\pi^{-}$ and $\eta'\pi^{-}$ final states}
\label{sec:4}
\vspace{-0.3cm}
\begin{figure}[tp!]
    \begin{center}
      \vspace{-0.2cm}
      \includegraphics[clip,trim= 40 140 40 120,width=1.0\linewidth]
      {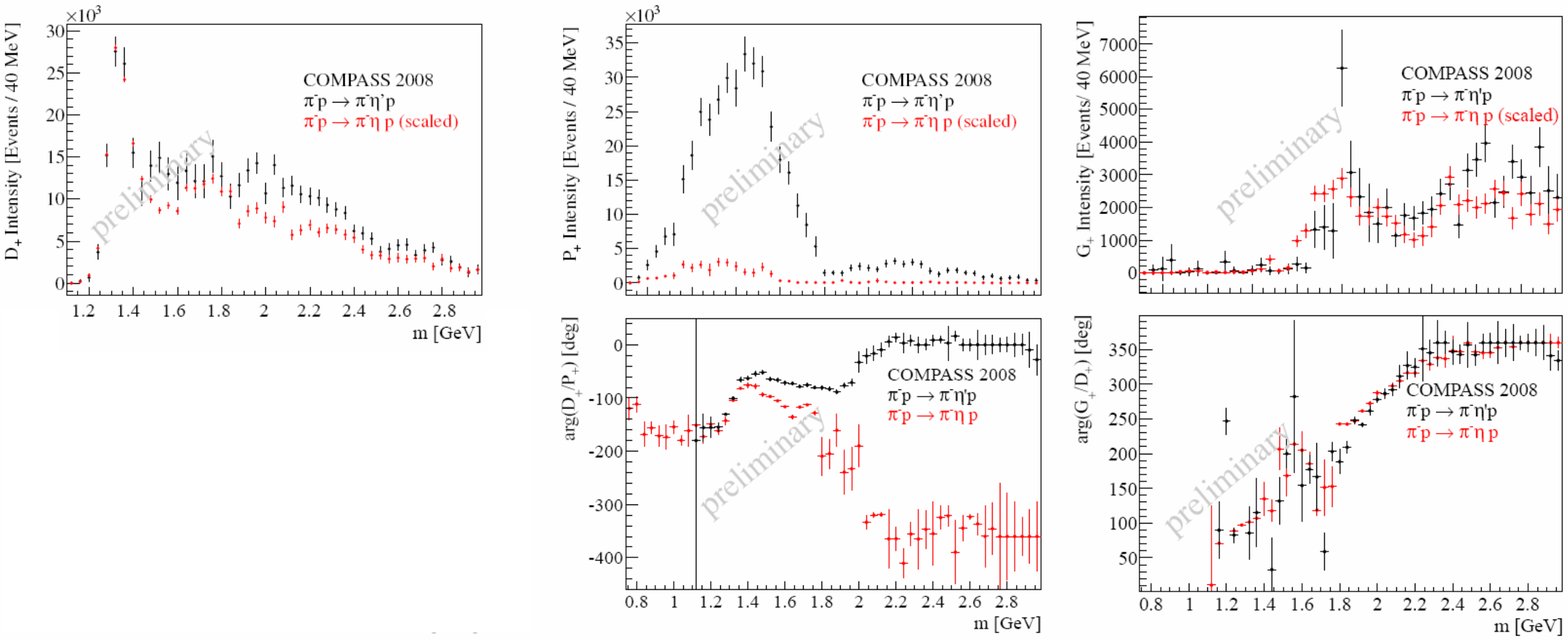}
      \vspace{-0.2cm}
      \caption{Mass-independent PWA results for $\eta\pi$ and $\eta'\pi$ final states for the 2008 data. 
	Shown are the fitted intensities for the $D+$ {\it (top/left)}, the spin-exotic $P+$ {\it (top/centre)} 
	and the $G+$ {\it (top/right)} and  waves. The relative phase differences with respect to the $D+$ 
	wave are shown for the $P+$ {\it (bottom/centre)} and the $G+$ {\it (bottom/right)} waves, respectively.} 	 
      \label{fig:etapi_etaprimepi}
      \vspace{-0.4cm}
    \end{center}
\end{figure}
Spin-exotic resonances including the $\pi_1(1600)$ were also reported in $\eta\pi^{-}$ and $\eta'\pi^{-}$ final states. 
The current status of these analyses at COMPASS is summarised in Fig.\,\ref{fig:etapi_etaprimepi}. The mass-independent 
PWA results compared for both decay channels are shown for the fitted intensities of the $D+$ (Fig.\,\ref{fig:etapi_etaprimepi}, 
top/left), the spin-exotic $P+$ (Fig.\,\ref{fig:etapi_etaprimepi}, top/centre), and the $G+$ (Fig.\,\ref{fig:etapi_etaprimepi}, 
top/right) waves as well as for the relative phase differences between the $P+$ (Fig.\,\ref{fig:etapi_etaprimepi}, bottom/centre) 
and the $G+$ wave (Fig.\,\ref{fig:etapi_etaprimepi}, bottom/right), both with respect to the $D+$ waves, respectively.  

The $D+$ and $G+$ waves appear nearly identical for the $\eta\pi^{-}$ and the $\eta'\pi^{-}$ cases after the corresponding phase 
space factors have been taken into account via normalisation. They show significant contributions that can be attributed to the 
$a_2(1320)$ ($D+$ wave) and the $a_4(2040)$ ($G+$ wave) resonances. The spin-exotic $P+$ wave, however, appears differently: The contribution 
in $\eta'\pi^{-}$ is much enhanced as compared to $\eta\pi^{-}$, inline with the observations by the VES collaboration~\cite{VES}. 
In particular the extracted phase motion (still to be taken with caution, as more systematic studies are ongoing) is found similar 
but stronger for $\eta\pi^{-}$ (Fig.\,\ref{fig:etapi_etaprimepi}, bottom/centre). The relative strengths of the $a_2(1320)$ and of 
the $a_4(2040)$ are consistent with theoretical predictions ($\eta - \eta'$ mixing). For a more detailed discussion, see \cite{tobi:2012}.
To summarise, COMPASS basically reproduces the results on $\eta\pi^{-}$ and $\eta'\pi^{-}$ as found by previous experiments.

\clearpage
\vspace{-0.5cm}
\subsection{Diffractively produced $K\bar{K}\pi\pi$ final states}
\label{sec:5}
\vspace{-0.3cm}
Another relevant and very interesting (as strangeness is involved) channel to search for the $\pi_1(1600)$ is the decay 
into $f_1\pi$ (in which also a spin-exotic $\pi_1(2000)$ candidate was reported), with $f_1 \rightarrow K\bar{K}\pi$, 
that is (among others) accessible at COMPASS in two different decay modes by analysing diffractively produced 
$K^{0}_{s}K^{\pm}\pi^{\mp}\pi^{-}$ final states.
\begin{figure}[tp!]
  \begin{minipage}[h]{.48\textwidth}
    \begin{center}
         \vspace{-0.3cm}
\resizebox{1.0\columnwidth}{!}{%
      \includegraphics[clip, trim= 12 10 12 20, width=1.0\linewidth]{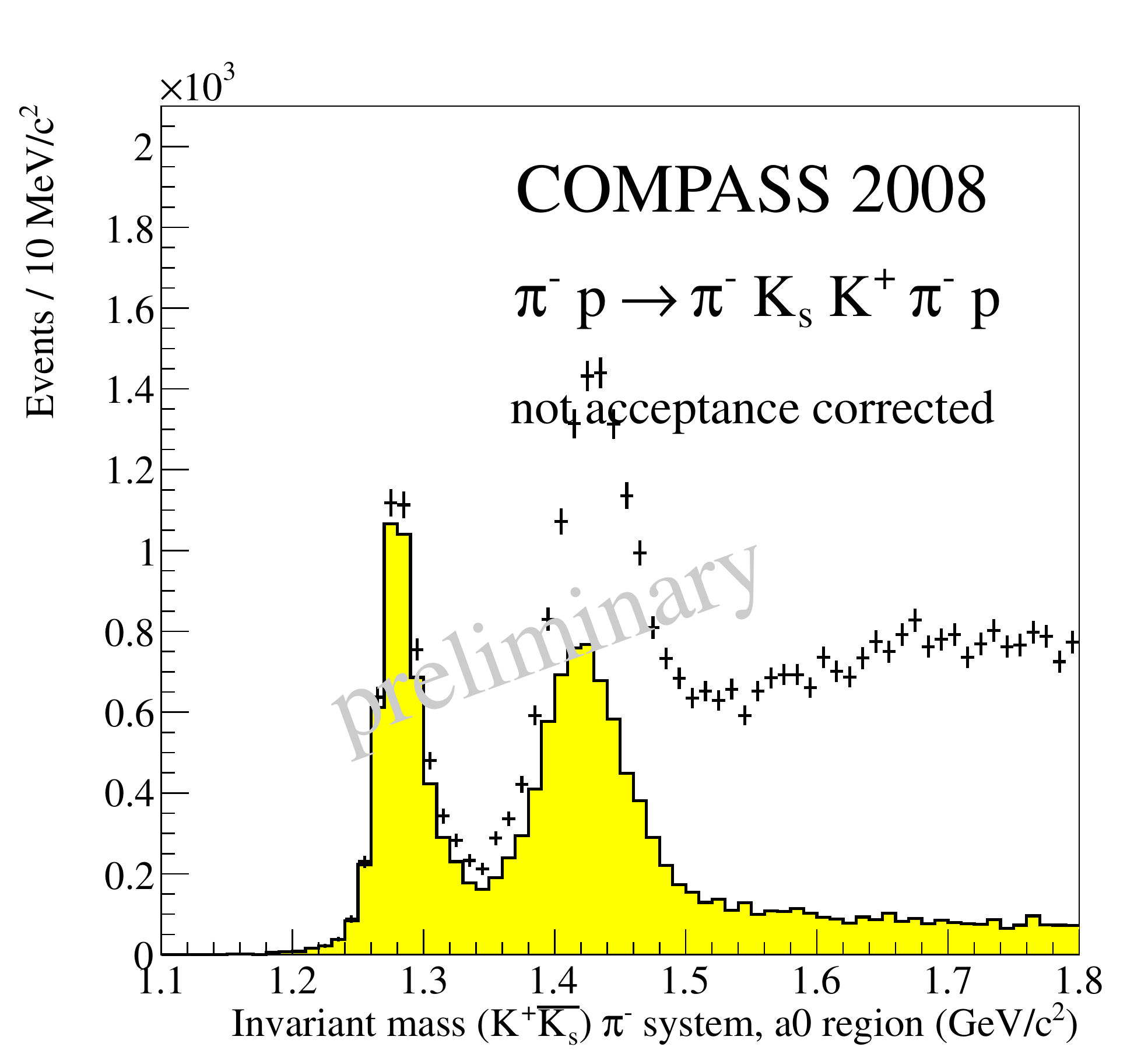}}
    \end{center}
  \end{minipage}
  \hfill
  \begin{minipage}[h]{.48\textwidth}
    \begin{center}
      \vspace{-0.3cm}
\resizebox{1.0\columnwidth}{!}{%
      \includegraphics[clip, trim = 440 80 25 155, width=1.0\linewidth]{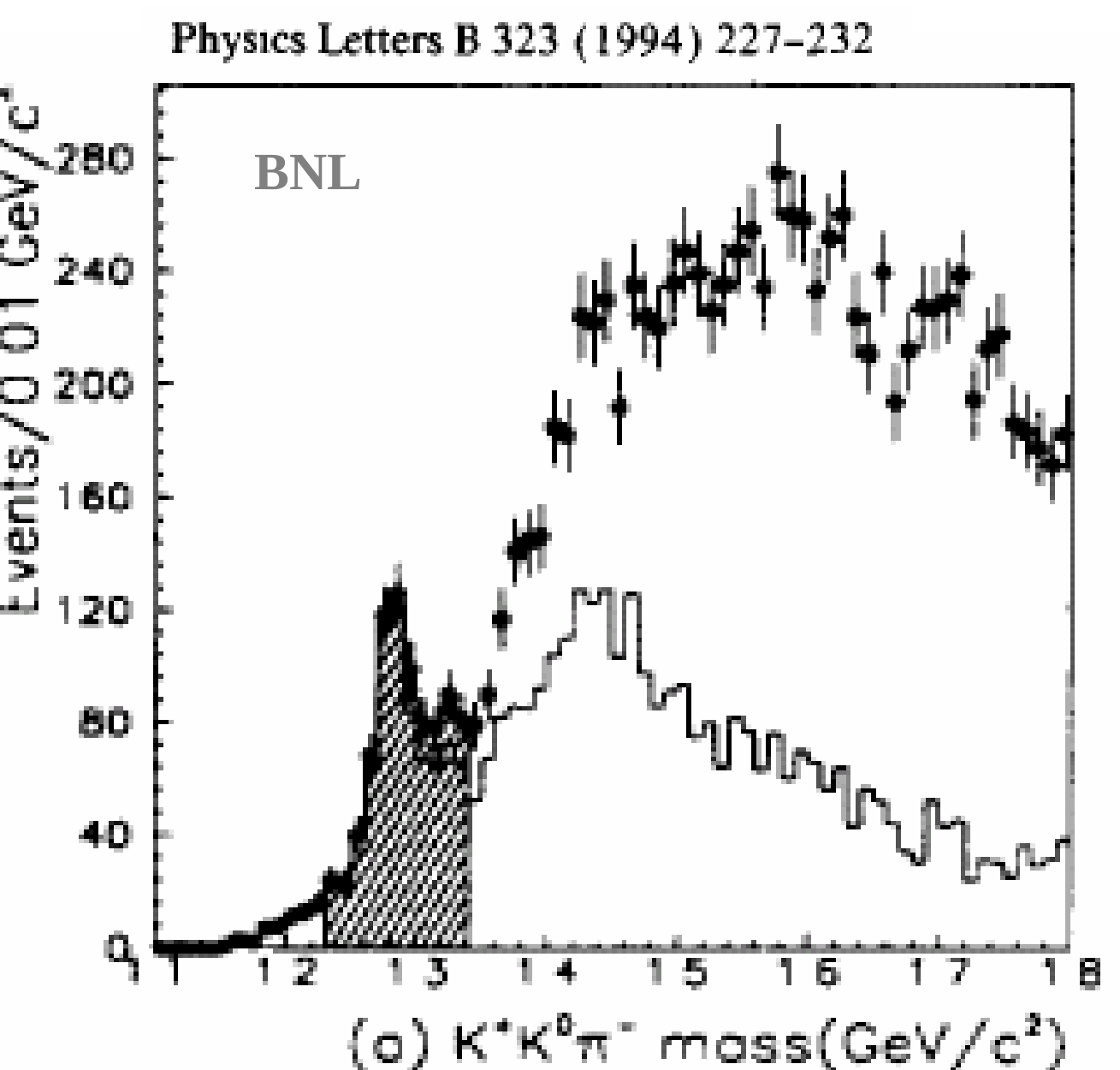}}
    \end{center}
  \end{minipage}
    \begin{center}
    \caption{{\it Left:} The $(K\bar{K}\pi)^{0}$ subsystem, showing clean $f_1(1285)$ and $f_1(1420)$ peaks 
before (dots) and after (line) an additional restriction of the $K\bar{K}$ mass to the $a_0(980)$ region. 
{\it Right:} Comparing the similar plot published by BNL/E818~\cite{JHLee:1994}, the COMPASS statistics exceeds 
the one analysed by E818 by a factor of 10 (or a factor of about 20, taking into account also the 
2009 data with negative pion beam). Not only the observed $f_1(1285)$ but also the $f_1(1420)$ are nearly 
background free as compared to BNL/E818~\cite{JHLee:1994}.}
\label{fig:BNL}
    \end{center}
\end{figure}
The resulting $(K\bar{K}\pi)^{0}$ subsystem is shown (exemplary for one case, the other looks similar) in 
Fig.\,\ref{fig:BNL}, left. The comparison to the corresponding plot published from the analysis carried out at BNL~\cite{JHLee:1994} 
(Fig.\,\ref{fig:BNL}, right) illustrates the quality of the 2008/09 hadron beam data. The COMPASS data feature 
clean, nearly background free $f_1(1285)$ and $f_1(1420)$ peaks. Even though an $\eta$ contribution cannot be excluded, 
a first mass-independent PWA indicate contributions from $\eta(1405)$ and $\eta(1295)$ to be minor, consistent with the 
observation by BNL/E852~\cite{Kuhn:2004}. Particularly interesting is the PWA of the $f_1(1420)\pi$ system, which to our best knowledge 
was never studied before and will be studied for the first time by COMPASS, for further details, see~\cite{bernhard_nerling:2011}. 

\subsection{Centrally produced $(\pi\pi)^{0}$ system from $pp$ reactions}
\label{sec:6}
\vspace{-0.3cm}
Apart of diffractive dissociation reactions, COMPASS is studying also central production. Part of the 2009 data set was 
taken with a 190\,GeV/$c^{2}$ proton beam scattered off the liquid hydrogen (proton) target, allowing for the 
search of exotic mesons and glue-ball candidates produced at central rapidities. A first event selection of 
the centrally produced $(\pi^{+}\pi^{-})$ system, on which a first model to describe the data in terms of 
partial-waves is being developed. First fits to the 2009 $pp$ data based on this ansatz deliver promising 
PWA results of the two-pseudoscalar final state consistent with those found by a previous experiment, namely the 
Omega Spectrometer at CERN (WA91)~\cite{WA91}. For the first results and details on the very preliminary analysis,
we refer the interested reader to a recent conference contribution~\cite{alex:2012}. 
The natural next step here will be the analysis of the $K^{+}K^{-}$ system, worth mentioning are also further 
channels available in the COMPASS 2009 $pp$ data such as $K_{\rm s}K_{\rm s}$, $\pi^{0}\pi^{0}$ and $\eta\eta$ --- 
COMPASS aims to shed new light also to this physics sector, improving the results by previous experiments in terms of 
statistics and resolution. 

\clearpage
\section{Test of Chiral Perturbation Theory and Primakoff reactions}
\label{sec:5}
Chiral Perturbation Theory (ChPT) effectively describes QCD at low energies, whereas the hadron dynamics are expressed
in terms of fundamental quantities ($m_{\pi}$, $f_{\pi}$ and higher-order low-energy constants) that have been adjusted 
to experimental measurements from {\it e.g.} $\pi\pi$ scattering, allowing for predictions of the fundamental quantities 
of the pion polarisability or the two-pion production cross-sections.

COMPASS performs such measurements in $\pi\gamma$ reactions (190\,GeV/$c^{2}$ $\pi^{-}$ beam scattering off 
the Coulomb field of a heavy ({\it e.g.} Pb) target. We briefly discuss here the first measurement of the cross-section 
$\pi^{-}\gamma \rightarrow \pi^{-}\pi^{+}\pi^{-}$ compared to a calculated ChPT prediction~\cite{kaiser:2008,kaiser:2010} 
recently published~\cite{Adolph:2012}, whereas data for a first polarisability determination via 
$\pi^{-}\gamma \rightarrow  \pi^{-}\gamma$ reactions has been taken in a short test run in 2009 (analysis ongoing) and in 
a dedicated data taking (already started) in 2012, both with a dedicated trigger for Primakoff events.
\begin{figure}[tp!]
    \begin{center}
      \vspace{-0.3cm}
      \includegraphics[clip,trim= 0 0 0 0,width=0.7\linewidth]
      {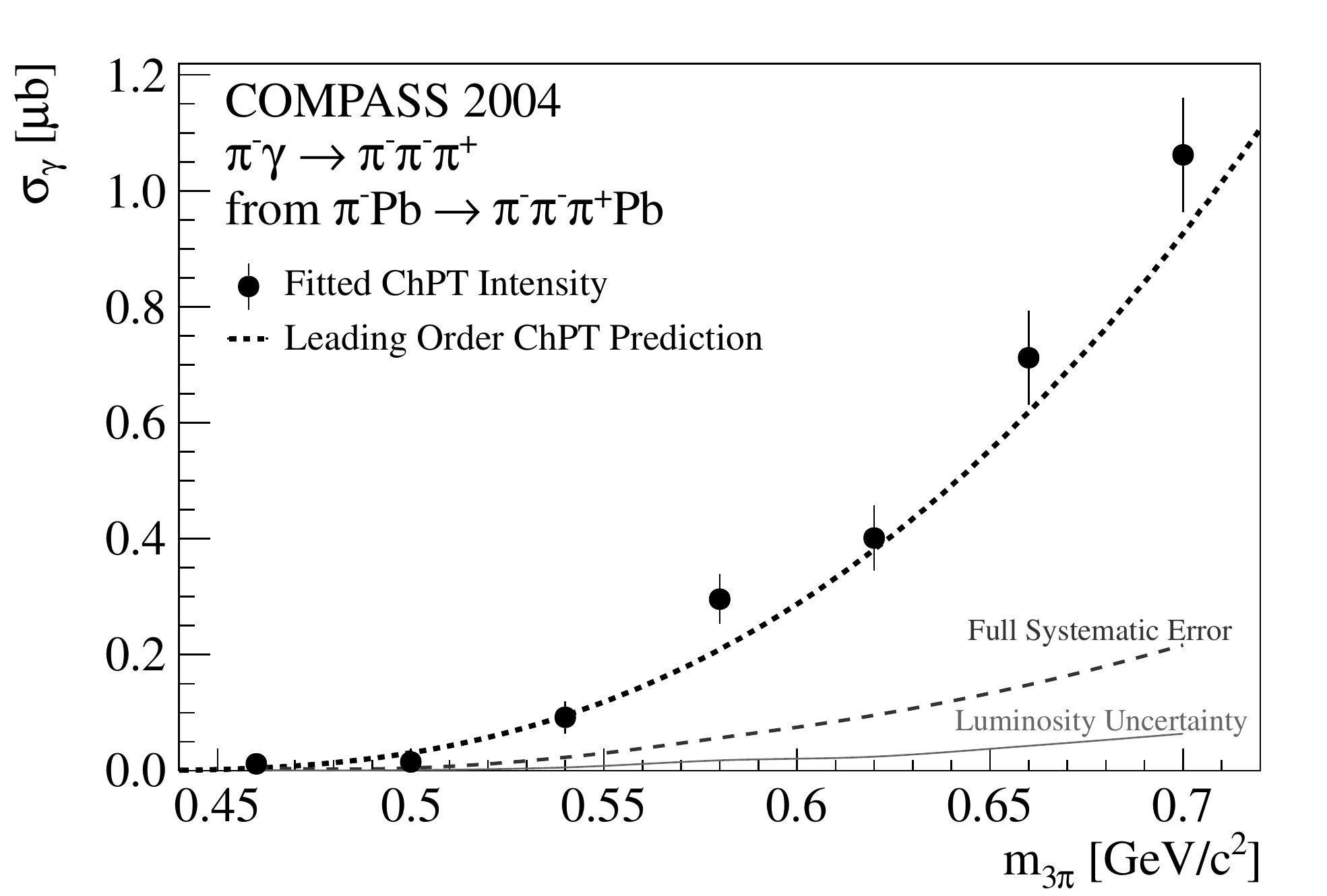}
      \vspace{-0.1cm}
      \caption{Measured absolute cross-section of $\pi^{-}\gamma \rightarrow \pi^{-}\pi^{+}\pi^{-}$ compared the corresponding 
	ChPT prediction~\cite{Adolph:2012}.}
      \label{fig:3p_ChPT}
      \vspace{-0.4cm}
    \end{center}
\end{figure}
A PWA has been performed on the 2004 pilot run data in 40\,MeV/$c^{2}$ mass bins for the momentum transfer range $t' < 10^{-3}$\,GeV$^{2}$/$c^{2}$ 
restricted to $0.44 < m_{3\pi} < 0.72$ GeV/$c^{2}$, using only the chiral amplitude for Primakoff production, means $M$=$1$ contribution. 
The cross-section is converted from the total intensity obtained by the fit to the experimental data (Pb target thickness and incoming 
beam flux are well determined). The final measured absolute cross-section of $\pi^{-}\gamma \rightarrow \pi^{-}\pi^{+}\pi^{-}$ derived from 
it is shown in Fig.\,\ref{fig:3p_ChPT} (data points) and compared to the corresponding ChPT calculation~\cite{kaiser:2008} (dotted line).
The data is found in good agreement with the theoretical prediction; for a more detailed discussion, see~\cite{Adolph:2012,Steffi:2012}.


\section{Summary \& Outlook}
\label{sec:6}
The status of various analyses of the COMPASS hadron data from the 2004 pilot run and the dedicated data taking campaigns in 2008/09
have been reported and discussed. Inline with the priorities, the focus has been given to the hadron spectroscopy programme in terms of 
PWA results in the context of the search for spin-exotic mesons. Especially the status of the search for the $\pi_1(1600)$ in the 2008 
COMPASS hadron beam data has been discussed, confirming the $(3\pi)^{-}$ PWA results published from the 2004 data (3 charged pion final states)
in two different decay modes (charged and neutral) on the 2008 proton target data, and extending the search to further decay channels, such 
as $\eta\pi$, $\eta'\pi$ and $f_1\pi$. Even though not relevant for the search for spin-exotic mesons, the first COMPASS PWA results on kaon 
diffraction should be mentioned~\cite{Promme:2011}, and also further physics measurements of interest are carried out at COMPASS, like 
{\it e.g.} the measurement of OZI violation in $\omega$ versus $\Phi$ production, showing a OZI violation of about a factor of 3 to 4 
(depending on $x_F$), these results are reported in a dedicated contribution at this conference~\cite{bernhard:2012}. 

To summarise, the COMPASS hadron data unique in terms of statistics for many channels are analysed in terms of various physics topics of 
relevance for and awaited by the community. Even though the first results are obtained, most of the physics results reported here need further
systematic studies before they will be published soon.  

\end{document}